\documentstyle[12pt]{article}
\newskip\humongous \humongous=0pt plus 1000pt minus 1000pt
\def\caja{\mathsurround=0pt}

\newif\ifdtup
\def\panorama{\global\dtuptrue \openup1\jot \caja
	\everycr{\noalign{\ifdtup \global\dtupfalse
	\vskip-\lineskiplimit \vskip\normallineskiplimit
	\else \penalty\interdisplaylinepenalty \fi}}}
\def\eqalignno#1{\panorama \tabskip=\humongous
	\halign to\displaywidth{\hfil$\displaystyle{##}$
	\tabskip=0pt&$\displaystyle{{}##}$\hfil
	\tabskip=\humongous&\llap{$##$}\tabskip=0pt
	\crcr#1\crcr}}
\def\tr{\mathop{\rm tr}}

\def\car{{\cal{R}}}

\def\wtp{{\widetilde{\partial}}}
\def\bb{\bf{B}}
\def\bc{\bf{C}}
\def\bbf{\bf{F}}

\def\wbr{{\widehat{\bf{R}}}}
\def\whr{\widehat{R}}

\def\ref#1{$^{#1)}$}
\begin{document}
\begin{titlepage}
\begin{center}
December 10,1992     \hfill    LBL-33249 \\
                     \hfill    UCB-PTH-92/41 \\

\vskip .5in

{\large \bf Differential Calculus on Quantum Spaces and Quantum Groups}
\footnote{This work was supported in part by the Director, Office of
Energy Research, Office of High Energy and Nuclear Physics, Division of
High Energy Physics of the U.S. Department of Energy under Contract
DE-AC03-76SF00098 and in part by the National Science Foundation under
grant PHY90-21139.}

\vskip .5in

Bruno Zumino\\[.5in]

{\em  Department of Physics\\
      University of California\\
      and\\
      Theoretical Physics Group\\
      Physics Division\\
      Lawrence Berkeley Laboratory\\
      1 Cyclotron Road\\
      Berkeley, California 94720}
\end{center}

\vskip .5in

\begin{abstract}
A review of recent developments in the quantum differential calculus. The
quantum group $GL_q(n)$ is treated by considering it as a particular quantum
space. Functions on $SL_q(n)$ are defined as a subclass of functions on
$GL_q(n)$.
The case of $SO_q(n)$ is also briefly considered. These notes cover part of a
lecture given at the XIX International Conference on Group Theoretic Methods in
Physics, Salamanca, Spain 1992.
\end{abstract}
\end{titlepage}
\noindent{\bf 1. Introduction}
\vskip 9pt

In this lecture I shall describe some recent developments in the theory of
differential calculi on quantum spaces and quantum groups. The general theory
is due to Woronowicz [1] and a number of interesting papers (see [2, 3]) have
elucidated various aspects of it.
I shall emphasize techniques [4] which give explicit commutation relations and
which are hopefully suitable for future physical applications.
Many of the conventions and notations used here can be found in [5].
This basic paper also contains numerous references.
\vskip 9pt

\noindent{\bf 2. Differential Calculus on Quantum Planes}
\vskip 9pt

We consider basic variables $x^k$, for $k= 1,2, ... n$,
which satisfy commutation relations
$$
B^{k\ell}_{mn} x^m x^n =0\eqno(2.1)
$$
where the $B^{k\ell}_{mn}$ are numerical coefficients.
We assume that these commutation relations allow one to order in some standard
way an arbitrary monomial in the variables.
Functions $f, g$ etc. of the basic variables can be defined as formal power
series and form an associative algebra. We wish to define an exterior
differential $d$ satisfying the usual underformed properties such as linearity,
plus
$$
d^2=0, \eqno(2.2)
$$
the Leibniz rule on functions (zero-forms)
$$
d(fg) = (df)g +f\ dg, \eqno(2.3)
$$
and
$$
d(dx^kf)=- dx^k df.\eqno(2.4)
$$
In (2.4) $f$ is a function or a differential form.
In general the differentials $dx^k$ of the basic variables will not commute
with the variables.
Here we consider the case when the commutation relations between the
differentials and the variables are bilinear
$$
x^k(dx^\ell) = C^{k\ell}_{mn} (dx^m) x^n,\eqno(2.5)
$$
where $C^{k\ell}_{mn}$ are numerical coefficients.

One can introduce derivatives on functions
$$
\partial_k \equiv {\partial\over \partial x^k} , \quad (\partial_k x^\ell) =
\delta^\ell_k\eqno(2.6)
$$
in the standard way through
$$
df = dx^k \partial_kf.\eqno(2.7)
$$
In general the derivatives do not satisfy the simple Leibniz rule of
commutative algebra.
We have, for an arbitrary function $f$,
$$
\eqalignno{
d(x^kf) &= (dx^\ell \partial_\ell x^k) f+x^k dx^\ell \partial_\ell f\cr
&= (dx^k) f + C^{k\ell}_{mn} (dx^m) x^n\partial_\ell f. &(2.8)\cr}
$$
which can be written as a commutation relation between derivatives and
variables
$$
\partial_\ell x^k = \delta_\ell^k +C^{km}_{\ell n} x^n \partial_m.\eqno(2.9)
$$
Applying $d$ to (2.5) one obtains, from (2.4),
$$
dx^k dx^\ell =- C^{k\ell}_{mn} dx^m dx^n.\eqno(2.10)
$$

In [6,7] commutation relations between derivatives and differentials where also
given, in the form
$$
\partial_k (dx^\ell) - D^{\ell m}_{kn}(dx^n) \partial_m =0 \eqno(2.11)
$$
and among the derivatives in the form
$$
\partial_n \partial_m F^{mn}_{k\ell} =0.\eqno(2.12)
$$

The coefficients $B, C, D,$ and $F$ must satisfy certain consistency relations,
discussed in [6,7].
There it was shown that it must be
$$
B^{k\ell}_{rs} + B^{k\ell}_{mn} C^{mn}_{rs} = 0.\eqno(2.13)
$$
This equation can be written in standard tensor product notation as
$$
B_{12} (I_{12} + C_{12}) =0 \eqno(2.14)
$$
where $I$ is the unit matrix. One finds also that $D=C^{-1}$, i.e.
$$
D^{k\ell}_{mn} C^{mn}_{rs} = C^{k\ell}_{mn} D^{mn}_{rs} = \delta^k_r
\delta^\ell_s,\eqno(2.15)
$$
 a Yang-Baxter equations for $C$
$$
C_{12} C_{23}C_{12} = C_{23}C_{12}C_{23}\eqno(2.16)
$$
and an orthogonality relation analogous to (2.14)
$$
(I_{12} + C_{12}) F_{12} =0.\eqno(2.17)
$$
Finally, we have two mixed Yang-Baxter equations:
$$
B_{12} C_{23} C_{12} = C_{23}C_{12}B_{23}\eqno(2.18)
$$
and
$$
C_{12}C_{23}F_{12} = F_{23} C_{12}C_{23},\eqno(2.19)
$$
which are sufficient conditions for the consistency of the calculus.

The above consistency conditions (2.13-19) are obtained by combining the
various
commutation relations.  For instance, multiply (2.1) from the left with
$\partial_r$ and commute this derivative through to the right by using (2.9)
twice.
One finds two terms which must vanish separately, the first proportional to a
single $x$, the second proportional to a product of the type $xx\partial$. The
vanishing of the first term gives (2.13), the vanishing of the second term is
ensured by (2.18).
The other conditions are obtained in a similar manner.

In many concrete examples the matrices $B, C,$ and $F$
can be expressed [8] as functions of a single matrix $\whr$ which
satisfies the Yang-Baxter equation
$$
\whr_{12} \whr_{23} \whr_{12} = \whr_{23}\whr_{12}\whr_{23}\eqno(2.20)
$$
and a characteristic equation
$$
(\whr - \mu_1) (\whr - \mu_2) ... (\whr - \mu_m) = 0. \eqno(2.21)
$$
 We assume that the eigenvalues $\mu_1, \mu_2, ... \mu_m$
are distinct; they may have different multiplicities.
For any particular non vanishing eigenvalue $\mu_\alpha$, one can choose
$$
C=- {\whr\over \mu_\alpha}\eqno(2.22)
$$
and
$$
B = F = \mathop{\Pi}_{\beta\neq\alpha} (\whr - \mu_\beta).\eqno(2.23)
$$
Clearly (2.16) is true, the orthogonality relations
(2.14) and (2.17) are obviously satisfied and the Yang-Baxter equations (2.18)
and (2.19) are also valid, because (2.20) implies
$$
p(\whr_{12}) \whr_{23}\whr_{12} = \whr_{23}\whr_{12}\ p(\whr_{23})\eqno(2.24)
$$
and
$$
\whr_{12}\whr_{23}\ p(\whr_{12})= p(\whr_{23})\whr_{12}\whr_{23}\eqno(2.25)
$$
for any polynomial $p$(.).

Notice that now (2.10) becomes.
$$
\whr^{ij}_{k\ell} dx^k dx^\ell = \mu_\alpha dx^i dx^j.\eqno(2.26)
$$
Therefore
$$
F^{ij}_{k\ell} dx^k dx^\ell = \mathop{\Pi}_{\beta\neq\alpha} (\mu_\alpha -
\mu_\beta) dx^i dx^j,\eqno(2.27)
$$
where $\alpha$ is fixed. Multiply this equation by $\partial_j\partial_i$ from
the right.
The left hand side vanishes by (2.12), so we obtain
$$
0={\mathop{\Pi}_{\beta\neq\alpha}} (\mu_\alpha - \mu_\beta) dx^i dx^j
\partial_j\partial_i.\eqno(2.28)
$$
Since $\mu_\beta \neq \mu_\alpha$ for all $\beta \neq \alpha$ ($\alpha$ fixed),
$$
d^2= d\ dx^i \partial_i =- dx^i d\partial_i =- dx^i dx^j
\partial_j\partial_i\eqno(2.29)
$$
vanishes in agreement with (2.2).

If all eigenvalues are different from zero one can use the inverse matrix
$\whr^{-1}$ which also satisfies the Yang-Baxter equation.
This gives alternative consistent forms of the calculus based on
$$
C=- \mu_\alpha \whr^{-1}
\eqno(2.30)$$
and
$$
B=F= {\mathop{\Pi}_{\beta\neq\alpha}} (\whr^{-1} - \mu^{-1}_\beta)\eqno(2.31)
$$
for any given eigenvalue $\mu_\alpha$ of $\whr$.

The simplest example that fits into the present scheme is that of the quantum
hyperplane where $\whr$ is the $\whr$-matrix of $GL_q(n)$ which satisfies a
characteristic equation with the two eigenvalues $\mu_1=q, \mu_2 =- q^{-1}$.
Here one can choose the eigenvalue $\mu_2$, which gives $C=q\whr, \ B=F=\whr
-q$.  The resulting calculus has been discussed in detail in [6, 7] (the
alternative based on (2.30) and (2.31) was also treated there).
An equivalent formulation is given in [9]. If one chooses instead the
eigenvalue
$\mu_1$ one obtains $C=- q^{-1}\whr, \ B=F=\whr + q^{-1}$, so that
the commutation relations are now
$$
x_1x_2=-q \whr_{12} x_1x_2,\eqno(2.32)
$$
$$
x_1 dx_2 =- {1\over q} \whr_{12} dx_1 x_2\eqno(2.33)
$$
and
$$
dx_1 dx_2 = {1\over q} \whr_{12} dx_1 dx_2.\eqno(2.34)
$$
As $q\to 1, \whr_{12}$ tends to $P_{12}$, the permutation matrix
$$
(P_{12})^{k\ell}_{mn} = \delta^k_n\delta^\ell_m. \eqno(2.35)
$$
Therefore the commutation relations become
$$
\eqalignno{
x^kx^\ell &=- x^\ell x^k&(2.36)\cr
x^kdx^\ell &= - dx^\ell x^k&(2.37)\cr
dx^kdx^\ell&= dx^\ell dx^k.&(2.38)\cr}
$$
In this limit the variables $x^k$ are fermionic but the commutation relations
(2.37) involving variables and differentials differ by a sign from the standard
ones for a simply graded fermionic calculus.
This is a perfectly consistent alternative with double grading  which goes
together with the validity of (2.3) and (2.4) also for fermionic $x$'s, and
which is equivalent [10], in a well defined sense, to the more standard simply
graded fermionic calculus .
The $q$-deformation of the standard fermionic calculus is given in [7]. An
equivalent formulation was presented in [11].

Another interesting example is that of quantum euclidian space, in which case
one takes the $\whr$-matrix of $SO_q(N)$, which satisfies a characteristic
equation with the three eigenvalues $\mu_1=q, \mu_2=-q^{-1}$ and $\mu_3
=q^{1-N}$ .  If one chooses the eigenvalue $\mu_2$ and applies the general
formulas (2.22) and (2.23) for $\alpha =2$, one obtains the conventional
quantum
calculus on euclidian space.

In the next section we shall see that the quantum group $GL_q(n)$ itself can be
treated as a quantum plane and that the calculus on $GL_q(n)$ fits into the
present formulation with an $\whr$-matrix having three distinct eigenvalues.
\vskip 9pt
\noindent{\bf 3. Calculus on $GL_q(n)$}
\vskip 9pt

The defining representation of the quantum group
$GL_q(n) $ is given in terms of $n \times n$ matrices
$A$ whose matrix elements satisfy the commutation relations
$$\whr_{12} A_1A_2 = A_1A_2 \whr_{12}\eqno(3.1)
$$
where the $\whr$ matrix of $GL_q(n)$ is given in [5] as
$$
\whr^{ij}_{k\ell} = \delta^i_\ell \delta^j_k (1+(q-1) \delta^{ij})
 + \lambda \delta^i_k\delta^j_\ell \theta_{ji}.\eqno(3.2)
$$
Here
$$
\lambda = q - {1\over q}\eqno(3.3)
$$
and
$$
\theta_{ji} = \Bigg\{\matrix{1& j>i\cr
                             0& j\leq i}.
\eqno(3.4)
$$
We shall take $q$ to be a generic complex number, not too far from 1.
The quantum determinant $det_q A$ of the matrix $A$ is defined [5] as
$$
det_q A = \sum_\sigma (-q)^{\ell(\sigma)} A_{1\sigma_1} A_{2\sigma_2} ...
 A_{n\sigma_n},\eqno(3.5)
$$
where the sum is over all permutations $(\sigma_1, \sigma_2, ... \sigma_n)$ of
the integers $(1,2, ... n)$ and $\ell(\sigma)$ is the length (number of
inversions) of the permutation $\sigma$.  The quantum determinant of $A$
commutes with all elements of $A$ as a consequence of (3.1).  We assume that it
does not vanish, so that we can define, at least in a formal sense, the matrix
$A^{-1}$.

Let us consider the matrix elements of $A$ as basic coordinates on group space.
 With the notation $({i\atop j} )=\alpha$,
$({k\atop\ell} )=\beta$ etc., we can write $A^i_j = x_\alpha$ etc.
The commutation relations (3.1) can be written in a form similar
 to (2.1) if we introduce a ``large'' $\wbr$-matrix defined by
$$
\wbr_{(12)(34)} = {1\over q}\whr_{13}\whr_{24}.\eqno(3.6)
$$
$\whr$ satisfies the characteristic equation
$$
(\whr - q) (\whr + {1\over q})=0.\eqno(3.7)
$$
Since its eigenvalues are $q$ and $-q^{-1}$,  those of $\wbr$
are $\mu_1=q, \mu_2 =-q^{-1}$ and $\mu_3 = q^{-3}$
$$
(\wbr -q) (\wbr + {1\over q}) (\wbr - {1\over q^3} ) =0.\eqno(3.8)
$$

If we choose the eigenvalue $\mu_2$ and apply the formulas of the previous
section we obtain
$$
{\bc} = q{\wbr},\eqno(3.9)
$$
$$
{\bb} = {\bbf} = ({\wbr} -q)({\wbr}- {1\over q^3})\eqno(3.10)
$$
and we are led to the commutation relations
$$
\left[(\wbr -q)(\wbr -{1\over q^3})\right]_{\alpha\beta,
 \gamma\delta} x_\gamma x_\delta = 0,\eqno(3.11)
$$
$$
x_\alpha dx_\beta = q \wbr_{\alpha\beta, \gamma\delta}
 dx_\gamma x_\delta\eqno(3.12)
$$
and
$$
dx_\alpha dx_\beta =- q \wbr_{\alpha\beta,
 \gamma\delta} dx_\gamma dx_\delta.\eqno(3.13)
$$
It is not hard to check that (3.11) is equivalent to (3.1)
and that (3.12) and (3.13) can be written respectively as
$$
A_1 dA_2 = \whr_{12} dA_1 A_2 \whr_{12}\eqno(3.14)
$$
and
$$
dA_1 dA_2 =- \whr_{12} dA_1 dA_2 \whr_{12}.\eqno(3.15)
$$
This is the form given by Schirrmacher [12] and Sudbery [13, 14] to the
commutation relations of Maltsiniotis [15, 16] and Manin [17, 18] for the
calculus on $GL_q(n)$.  We see that they agree with the general formulation
of Sec. 2 for the calculus on a quantum plane (notice that
 the characteristic equation (3.8) is the same as for $SO_q(4) \sim
SL_q(2) \times SL_q(2)).$

It is convenient to introduce the numerical diagonal matrix
$$
D= diag (1, q^2, ... , q^{2(n-1)}).\eqno(3.16)
$$
This matrix satisfies a number of useful relations which are listed in [4, 5,
19].  In particular, for any $n \times n$ matrix $M$, it is
$$
{\tr} _1 (D^{-1}_1 \whr^{-1}_{12} M_2 \whr_{12})= \tr (D^{-1}M) I_2 \eqno(3.17)
$$
where $\tr_1$ is the trace with respect to the indices
relative to the first space in the tensor product and $I_2$
is the unit matrix in the second space.
Also
$$
{\tr} _1 (D^{-1}_1 \whr^{-1}_{12}) = q^{1-2n} I_2.\eqno(3.18)
$$
If $A$ satisfies (3.1) then
$$
D^{-1} A^t D(A^{-1})^t = (A^{-1})^t D^{-1} A^t D=I,\eqno(3.19)
$$
where $^t$ denotes transposition. It follows that, if the matrix elements of
$M$
commute with those of $A$, then
$$\tr (D^{-1} A^{-1} MA) = \tr (D^{-1}M).\eqno(3.20)
$$
For this reason, $\tr (D^{-1} M)$ is called the quantum invariant trace of $M$.

As we know from the previous section, one can introduce derivatives which,
 according to (2.9) and (2.12) satisfy now
$$
\partial_\alpha x_\beta = \delta_{\alpha\beta} + q \wbr_{\beta\delta,
 \alpha\gamma} x_\gamma \partial_\delta \eqno(3.21)
$$
and
$$
\partial_\beta\partial_\alpha \left[(\wbr -q)(\wbr -
 {1\over q^3})\right]_{\alpha\beta, \gamma\delta}=0.\eqno(3.22)
$$
These equations can be written in a form analogous to (3.1), (3.14) and (3.15).
Let us introduce a matrix of derivatives by $\widetilde{\partial}^i_j =
\partial_\alpha$ and $\partial^i_j =(D^{-1})^j_k\wtp^k_i$.  Using (3.17) and
(3.18) one finds
$$
\partial_1\whr_{12}^{-1}A_1=q^{1-2n} I_{12}+A_2\whr_{12}\partial_2\eqno(3.23)
$$
and
$$
\whr_{12} \partial_2\partial_1 = \partial_2\partial_1\whr_{12}.\eqno(3.24)
$$

Equations (3.1), (3.14) and (3.15) go into themselves under the left coaction
$A\to A'A$ and the right coaction $A\to AA'$ where $A'$
is a constant (i.e. $dA' =0$), $GL_q(n)$ matrix which satisfies (3.1).
Equations (3.23) and (3.24) also go into themselves if one transforms the
 derivative matrix respectively as $\partial \to
\partial (A')^{-1}$ and $\partial \to
(A')^{-1}\partial$ (the constancy of the matrix $A'$ implies that its matrix
elements commute with those of $dA$ and of $\partial$).

The Cartan-Maurer form
$$
\Omega = A^{-1} \ dA\eqno(3.25)
$$
is left-invariant and right-covariant i.e. $\Omega \to \Omega$
and $\Omega \to (A')^{-1}\Omega A'$ under the respective coactions
above.
The 1-form
$$
\xi =- q^{2n-1} \tr (D^{-1}\Omega)\eqno(3.26)
$$
is both left- and right-invariant, see (3.20).
$\Omega$ satisfies the following equations due to (3.1),
(3.14) and (3.15)
$$
\eqalignno{
\Omega_1 A_2 &= A_2 R^{-1}_{12} \Omega_1 R^{-1}_{21}&(3.27)\cr
\Omega_1dA_2 &+ dA_2 R^{-1}_{12} \Omega_1 R_{12} =0&(3.28)\cr
\Omega_1 R^{-1}_{21}\Omega_2 R_{21} &+ R^{-1}_{21} \Omega_2
 R^{-1}_{12}\Omega_1 =0.&(3.29)\cr}
$$
Here and in the followings we use the $R$-matrix of $GL_q(n)$, which is related
to the $\whr$-matrix used above by
$$
(R_{12})^{ij}_{k\ell} = (P_{12}\whr_{12})^{ij}_{k\ell} =
 (\whr_{12})^{ji}_{k\ell}.\eqno(3.30)
$$
Thus (3.1) becomes
$$
R_{12} A_1A_2 = A_2A_1 R_{12}.\eqno(3.31)
$$
{}From the properties of $D$ and the characteristic equation (3.7)
 one can show [4] that the above equations imply
$$
dA = \lambda^{-1}(\xi A - A\xi)\eqno(3.32)
$$
and
$$
d\Omega =- \Omega^2 = \lambda^{-1}(\xi \Omega + \Omega \xi ). \eqno(3.33)
$$
Thus, if $f$ is any form
$$
df = \lambda^{-1}[\xi, f]_\pm,\eqno(3.34)
$$
where [\ ,  ]$_\pm$ is a commutator for even degree forms, an anticommutator
for
odd degree forms ($\lambda$ is given in (3.3)).

The quantum determinant $det_qA$ of the matrix $A$ is a zero form.
We know that it commutes with all elements of $A$.
The above equations imply that
$$
\Omega (det_q A) = q^{-2} (det_q A) \Omega\eqno(3.35)
$$
and
$$
d(det_q A) =- q^{-1} (det_q A) \xi =- q\xi (det_q A).\eqno(3.36)
$$
A consequence of these equations is that both $d\xi$ and $\xi^2$ vanish.
The elements of $\Omega$ form a linearly independent basis for 1-forms, and we
shall use them instead of the elements of $dA$ from
now on.
\vskip 9pt
\noindent{\bf 4. Inner Derivations and Lie Derivatives for $GL_q(n)$.}
\vskip 9pt

Following [4], we now introduce the inner derivation, which we take to be a
left
action mapping $k$-forms to $(k-1)$-forms.  Its action on the $n^{2}$ elements
of $A$ and $\Omega$ is given by introducing $n^{2}$ vector fields
$X^{i}{}_{j}$,
and the associated $n^{2}$ inner derivations are the entries in the matrix
$i_{X}$ whose elements are
$$
(i_{X})^{i}{}_{j}=i_{X^{i}{}_{j}}.\eqno(4.1)
$$
$i_{X}$ must act on 0- and 1-forms in a way preserving the commutation
relations
(3.31) and (3.27-29); the appropriate actions are
$$
\eqalignno{
i_{X_{1}}A_{2}&=A_{2}R_{21}i_{X_{1}}R_{12},&(4.2)\cr
R_{21}i_{X_{1}}R_{12}\Omega _{2} +\Omega_{2} R_{21}i_{X_{1}}R_{12} &=\frac{1-
R_{21}R_{12}}{\lambda}.&(4.3)\cr}
$$
These two equations imply that when evaluated on 0- and 1-forms,
$$
(i_{X}f)=0, \ (i_{X_{1}}\Omega_{2})=-q^{1-2n}D_{2}P_{12},\eqno(4.4)
$$
where $f$ is any function of the elements of $A$.  Equation (4.3) gives
$$
i_{X} \xi +\xi i_{X}=I.\eqno(4.5)
$$
Notice that by using the characteristic equation $\frac{1-R_{21}R_{12}}
{\lambda}$ could be replaced by $-\hat{R}_{12}$.  On $det_{q}A$, the
inner derivation acts as
$$
i_{X}(det_{q}A)=q^{2}(det_{q}A)i_{X}.\eqno(4.6)
$$
The commutation relations between the inner derivation matrices are
$$
R_{12}^{-1}i_{X_{1}}R_{12}\, i_{X_{2}}+i_{X_{2}}R_{21}i_{X_{1}}R_{12}=0.
\eqno(4.7)
$$
It is easy to see
that $i_{X}$ is left-invariant and right-covariant
under the respective coactions on $A$.

We may now introduce the Lie derivative matrix $L_{X}$ in the same way as in
the classical theory, i.e. a left action taking $k$-forms to $k$-forms given by
$$
L_{X}\equiv i_{X}d+di_{X},\eqno(4.8)
$$
where $L_{X}$ is a matrix with elements $L_{X^i_{j}}$ which by definition
transforms in the same way as $i_{X}$ does.  The equations already given for
$d$ and $i_{X}$ imply the following relations involving $L_{X}$:
$$
\eqalignno{
L_{X}d &= dL_{X},&(4.9) \cr
R_{21}L_{X_{1}}R_{12}\, i_{X_{2}}-i_{X_{2}} R_{21} L_{X_{1}} R_{12}&=
\lambda^{-1} (R_{21} R_{12} i_{X_{2}}-i_{X_{2}} R_{21}R_{12}), &(4.10)\cr
R_{21} L_{X_{1}} R_{12} L_{X_{2}} - L_{X_{2}} R_{21} L_{X_{1}} R_{12}&=
\lambda^{-1}(R_{21}R_{12}L_{X_{2}} -L_{X_{2}}R_{21}R_{12}),&(4.11)\cr
L_{X_{1}}A_{2}&=A_{2}R_{21}L_{X_{1}} R_{12}+A_{2}(\frac{1-R_{21}R_{12}}
{\lambda}),\cr
&&(4.12)\cr
R_{21}L_{X_{1}}R_{12}\Omega_{2}-\Omega_{2}R_{21}L_{X_{1}}R_{12}&=
\lambda^{-1}(R_{21}R_{12}\Omega_{2}-\Omega_{2}R_{21}R_{12}),&(4.13)\cr
L_{X}\xi&=\xi L_{X},&(4.14)\cr}
$$
and for the determinant,
$$
L_{X}(det_{q}A)=q^{2}(det_{q}A)L_{X}-q(det_{q}A).\eqno(4.15)
$$
Many of these relations take a much simpler form if we introduce the
Lie derivative valued operator $Y$ given by
$$
Y=1-\lambda L_{X},\eqno(4.16)
$$
which, of course, has the same transformation properties as $L_{X}$.
Using this, we obtain
$$
\eqalignno{
Yd &=dY,&(4.17) \cr
R_{21}Y_{1}R_{12}\, i_{X_{2}}&=i_{X_{2}}R_{21}Y_{1}R_{12}, &(4.18)\cr
R_{21}Y_{1}R_{12}Y_{2}&=Y_{2}R_{21}Y_{1}R_{12},&(4.19)\cr
Y_{1}A_{2}&=A_{2}R_{21}Y_{1}R_{12},&(4.20)\cr
R_{21}Y_{1}R_{12}\Omega_{2}&=\Omega_{2}R_{21}Y_{1}R_{12}, &(4.21)\cr
Y\xi&=\xi Y,&(4.22)
\cr}
$$
and
$$
Y(det_{q}A)=q^{2}(det_{q}A)Y.\eqno(4.23)
$$
A matrix satisfying (4.19) was introduced on several occasions in the
literature
(see [20, 21]) and is often called $L$ instead of $Y$; (4.19) is often called
the ``reflection equation''.

The matrix $Y$ is invertible, at least in a formal sense. It is also possible
[4, 22] to define a quantum determinant Det $Y$ which commutes with the
elements of $Y$.  Perhaps the simplest way to introduce it is to observe that
the matrix $AY$ satisfies the same commutation relations (3.1, 31) as the
matrix
$A$ itself as can be seen  using (3.1, 31), (4.19) and (4.20).  We can define
$$
Det Y = q^{n(n-1)} [det_q \ A]^{-1} [det_q (AY)],\eqno(4.24)
$$
where the right hand side involves only the standard quantum determinant.
Alternatively, if we observe that $YA^{-1}$ satisfies the same commutation
relations as $A^{-1}$, we can write
$$
Det Y = q^{n(n-1)} [det_{q^{-1}} (YA^{-1})] [det_{q^{-1}} A^{-1}],\eqno(4.25)
$$
which gives an equivalent result.
This determinant is invariant under transformations of $Y$  (i.e. $Y \mapsto
Y$ for $A \mapsto AA^{\prime}$ and $Y \mapsto (A{^\prime) ^{-1}}YA^{\prime}$
for
$A \mapsto A^{\prime}A$, with $Y$ and $A^{\prime}$ having commuting elements),
and satisfies the following relations:
$$
\eqalignno{
d (Det\, Y)&=(Det\, Y)d,&(4.26) \cr
(Det\, Y)i_{X}&=i_{X}(Det\, Y), &(4.27)\cr
(Det\, Y)A&=q^{2}A(Det\, Y), &(4.28)\cr
(Det\, Y)\Omega &= \Omega (Det\, Y), &(4.29)\cr
(Det\, Y)\xi &=\xi (Det\, Y), &(4.30)\cr}
$$
and
$$
(Det Y)(det_{q}A)=q^{2n}(det_{q}A)(Det\, Y).\eqno(4.31)
$$
The above equations for $Det\, Y$ suggest the definition of an operator
$H_{0}$ as
$$
Det \, Y
 \equiv q^{2H_{0}}. \eqno(4.32)
$$
$H_{0}$ commutes with $Y$, $d$, $i_{X}$, $\Omega$, and $\xi$, and satisfies
$$
[H_{0},A]=A,\ \  [H_{0},det_{q}A]=n(det_{q}A). \eqno(4.33)
$$
\vskip 9pt
\noindent{\bf 5. Calculus on the Quantum Group $SL_q(n)$}
\vskip 9pt

There seems to be an obvious way to specify the calculus on the
 quantum group $SL_{q}(N)$:
take the matrix $A$ and set its  quantum determinant to unity.
However,  although $det_{q}A$ commutes with
 the elements of $A$, it does {\em not} commute with such quantities
as $\Omega$ and $Y$.  Therefore, instead of imposing $det_{q}A=1$, we {\em
define} matrices $T$ as
$$
T=(det_{q}A)^{-1/n}A. \eqno(5.1)
$$
With $det_{q}T$ defined as in (3.5), the centrality of $det_{q}A$
automatically gives $T$ determinant unity.
This matrix $T$ is what we
identify as an element of the defining representation of $SL_{q}(N)$, since
it also satisfies (3.1) with $A$ replaced by $T$.  As we will
see in the next section, it becomes convenient to introduce the matrix
$$
{\cal R}_{12} = q^{-1/n}R_{12},\eqno(5.2)
$$
which we identify as the R-matrix for $SL_{q}(N)$.  Thus, we shall write
(3.1) as
$$
{\cal R}_{12}T_{1}T_{2}=T_{2}T_{1}{\cal R}_{12}.\eqno(5.3)
$$

The exterior derivative on $SL_{q}(n)$ can be taken to be the same as that
introduced on $GL_{q}(n)$; this is because $T$ is a function of the  elements
of $A$, so its differentials are given by
$$
dT=\lambda^{-1}[\xi ,T].\eqno(5.4)
$$
Note that this implies that the Cartan-Maurer form $\tilde{\Omega}$ for
$SL_{q}(n)$ is given by
$$
\tilde{\Omega}\equiv T^{-1}dT=q^{2/n}\Omega +q \
[1/n]_{q}\xi,
\footnote{This relation implies that the matrix of differential forms
introduced in [19] is equal to $-q^{2n-1}\Omega$.}\eqno(5.5)
$$
where
$$
[x]_q = {1-q^{2x}\over 1-q^2}.\eqno(5.6)
$$
In the classical limit $q \rightarrow 1$, $\tilde{\Omega}$ is
traceless, giving the appropriate reduction from $n^{2}$ to $n^{2}-1$
independent elements in the Cartan-Maurer matrix 1-form for $SL(n)$.

We have thus found a way to set the determinant of our $SL_{q}(n)$ matrices to
unity; for the calculus on the group, we must do something similar, namely
impose a constraint so that the number of independent differential operators is
reduced from $n^{2}$ to $n^{2}-1$.  In a way, we have already done this,
because
(4.33) and (5.1) together imply
$$
[H_{0},T]=0,\eqno(5.7)
$$
so that $H_{0}$ commutes with everything of interest in $SL_{q}(n)$, i.e.
matrices, forms, exterior derivative, etc.  Thus, within the context of
$SL_{q}(n)$, $H_{0}$ is irrelevant, reducing the number of generators
from $n^{2}$ to $n^{2}-1$, as desired.  Explicitly, this restriction is
accomplished by defining a new Lie derivative valued operator $Z$ by
$$
Z\equiv q^{-2H_{0}/n}Y.\footnote{When restricted to acting on 0-forms,
this operator is identical to the operator $Y$ in [19].}\eqno(5.8)
$$
Note that the determinant of $Z$, computed using e.g. (4.24), is unity.  This
is equivalent to the introduction of a set of $n^{2}$ ``vector fields''
$V^{i}{}_{j}$ through $Z=1-\lambda L_{V}$, so that
$$
L_{V}=L_{X}+q^{-1}[H_{0}/n]_{q^{-1}}-q^{-1}\lambda
L_{X}[H_{0}/n]_{q^{-1}}.\eqno(5.9)
$$
The fact that $Det\, Z=1$ implies that only $n^{2}-1$ of the elements of
$L_{V}$
are actually independent, which is precisely what we require for $SL_{q}(n)$.
In the classical limit, $H_{0}=-tr(L_{X})$,  so $L_{V}$ becomes traceless;
thus,
$V$ contains only $n^{2}-1$ linearly independent vector fields, as we would
expect.

Now that we have obtained all these quantities, we want to find the various
relations they satisfy.
The commutation relations between $\Omega$ and $T$ are given by
$$
\Omega_{1}T_{2}=q^{2/n}T_{2}R_{12}^{-1}\Omega_{1}R_{21}^{-1}
=T_{2}{\cal R}_{12}^{-1}\Omega_{1}{\cal R}_{21}^{-1}.\eqno(5.10)
$$
Here we see the appearance of ${\cal R}_{12}$, as promised.
$\Omega$ remains
unchanged, so  (3.29) is still valid: it does not have ${\cal R}
_{12}$ in place of $R_{12}$.  $L_{V}$ satisfies
$$
{\cal R}_{21}L_{V_{1}}{\cal R}_{12}L_{V_{2}}-L_{V_{2}}{\cal
R}_{21}L_{V_{1}}{\cal R}_{12}=\lambda^{-1}({\cal R}_{21}{\cal
R}_{12}L_{V_{2}}-L_{V_{2}}{\cal R}_{21}{\cal R}_{12}),\eqno(5.11)
$$
The actions of the various operators on the 0- and 1-forms of $SL_{q}(n)$
are given by
$$
L_{V_{1}}T_{2}=T_{2}{\cal R}_{21}L_{V_{1}}{\cal
R}_{12}+T_{2}(\frac{1-{\cal R}_{21}{\cal R}_{12}}{\lambda}), \eqno(5.12)
$$
and
$$
{\cal R}_{21}L_{V_{1}}{\cal R}_{12}\Omega _{2}-\Omega_{2}{\cal R}_{21}
L_{V_{1}}{\cal R}_{12}=\lambda^{-1}({\cal R}_{21}{\cal R}_{12}\Omega_{2}
-\Omega_{2}{\cal R}_{21}{\cal R}_{12}).\eqno(5.13)
$$
As a consequence, $\xi$ satisfies
$$
L_{V}\xi = \xi L_{V}.\eqno(5.14)
$$
The relations for $Z$ corresponding to (4.17-22) are
$$
\eqalignno{
\car_{21} Z_1 \car_{12} Z_2 &= Z_2 \car_{21} Z_1 \car_{12},&(5.14)\cr
Z_1T_2 &= T_2\car_{21}Z_1\car_{12},&(5.16)\cr}
$$
$$
\car_{21} Z_1\car_{12} \Omega_2 = \Omega_2\car_{21}Z_1\car_{12}\eqno(5.17)
$$
and
$$
Z\xi = \xi Z.\eqno(5.18)
$$
Notice that the invariant form constructed with the Cartan-Maurer
 form $\widetilde{\Omega}$ is
$$
\widetilde{\xi} = tr \ D^{-1}\widetilde{\Omega} =- q^{1-2n} (q^{2/n} - q^{2n}
\left[{1\over n}\right]_q [n]_{1/q})\xi .\eqno(5.19)
$$
It vanishes as $q\to 1$ as it should.

\vskip 9pt
\noindent{\bf 6. Conclusion}
\vskip 9pt

In Sec. 3 we have seen that the differential calculus for $GL_q(n)$ is a
special
case of the differential calculus on quantum planes. We chose there the $\wbr$
version of the $GL_q(n)$ calculus, the $\wbr^{-1}$ version could be developed
in
a similar way.

In Sec. 5 we derived the differential calculus on $SL_q(n)$ by defining the
functions on $SL_q(n)$ as a subclass of functions on $GL_q(n)$.
While for $GL_q(n)$ there are $n^2$ independent Lie derivative operators $Y$
and $n^2$ independent 1-forms $\Omega$, for $SL_q(n)$ the number of Lie
derivatives is reduced to $n^2-1$  by the relation  $Det Z=1$.
However, the number of Cartan Maurer 1-forms is still $n^2$.
One of them is the invariant form $\xi$ which generates the differentiation
through (5.4) and (3.32), and which has no classical analogue.
This is related to the fact that, in spite of the restriction $Det Z=1$, it is
not possible to find $n^2-1$ Lie derivatives which satisfy a bicovariant
deformed Lie algebra with only quadratic relations.
The $n^2$ elements of $Z$ are of this type and the relation $Det Z=1$ is
consistent with the commutation relations, but it is a polynomial relation.
If one drops the requirement of bicovariance, for $SL_q(2)$ there exist a right
invariant and also a left invariant calculus with $n^2-1=3$ Lie derivatives
satisfying quadratic commutation relations.  However, this seems to be a
special
property of $n=2$.  For higher $n$ no such calculi with $n^2-1$ Lie derivatives
are known, even if one drops the requirement of bicovariance.

An important lesson one can derive from the developments of the previous
 sections is that it is very useful to consider
the larger algebra which has as generators for $GL_q(n)$ the matrix
 elements of $A$ and of $Y$ together (or of
$T$ and $Z$ for $SL_q(n))$, their commutation relations being given by (3.31)
and (4.19, 20).  While the functions on the group form a Hopf algebra and the
enveloping algebra of the $Y$ is also a Hopf algebra, the $A, Y$ larger algebra
is not a Hopf algebra; still, it contains all the necessary information.
This point of view, which allows multiplication of elements of $A$ with
elements
of $Y$, leads to the simple definition of the $Det Y$ given in (4.24, 25).

The consideration of the larger $A, Y$ (or $T, Y$) algebra is useful for other
quantum groups as well.  For instance, for $SO_q(N)$, there is an orthogonality
relation for the $T$ matrices
$$
T^t CT =C, \eqno(6.1)
$$
where the metric matrix $C$ is defined in [5].
For this quantum group, the matrix product $q^{N-1} TZ$ satisfies all
 relations for
a quantum orthogonal matrix, including (6.1).
Indeed, one can verify that
$$
q^{2(N-1)} (TZ)^t C(TZ)=C\eqno(6.2)
$$
gives rise to the correct relations for the $Z$ matrix of $SO_q(N)$, i.e.
[22, 24]
$$
q^{N-1} C_{k\ell} Z^\ell_m R^{mk}_{in} Z^n_{j} = C_{ij},\eqno(6.3)
$$
where $R^{mk}_{in}$ is here the $R$-matrix of $SO_q(N)$ given in [5].
This can be easily seen using (6.1), the relation
$$
C_{ij} R^{ji}_{k\ell} = q^{1-N}C_{k\ell}.\eqno(6.4)
$$
and the $Z-T$ commutation relation, which for $SO_q(N)$, is still
$$Z_1T_2 = T_2 R_{21}Z_2 R_{12}.\eqno(6.5)
$$

For $SO_q(N)$ the situation described earlier for $SL_q(n)$ is even more
extreme.  The number of independent Lie derivatives is reduced from $n^2$ to
$n(n-1)/2$ by the  polynomial relations (6.3).
However, the number of independent Cartan-Maurer 1-forms is still $n^2$.
Of these, one is the invariant 1-form $\xi$ which plays a special role
analogous
to that for $GL_q(n)$ or $SL_q(n)$, but now there are\\
 $n^2-1-n(n-1)/2 = n(n+1)/2-1$ additional
1-forms which cannot be eliminated in the bicovariant calculus.
Only as $q\to 1$ these 1-forms vanish [23] in the combination
$\widetilde{\Omega} = T^{-1} dT$.  Except for the case of $GL_q(n)$ (and for
the
nonbicovariant calculi on $SL_q(2))$, the introduction of all the additional
1-forms seems unavoidable.  The elegant commutation relations for Lie
derivatives involving only quadratic (and linear) terms seems possible only at
the price of introducing more of them than in the classical $q=1$ case and then
restricting their number by means of polynomial relations.
These facts are worth emphasizing, since they are mostly ignored in the
literature.

\newpage
\noindent{\bf References}
\begin{enumerate}
\item S.L. Woronowicz, Commun. Math. Phys. {\bf 122} 125 (1989).
\item D. Bernard, Prog.  Theoretical Phys. Supp. {\bf 102} 49 (1990).
\item B. Jur\v{c}o, Lett. Math. Phys. {\bf 22} 177 (1991).
\item P. Schupp, P. Watts, and B. Zumino,  Lett. Math. Phys. {\bf 25} 139
(1992).
\item N. Yu. Reshetikhin, L. A. Takhtadzhyan, and L. D. Faddeev,
Leningrad Math. J. {\bf 1} 193 (1990).
\item J. Wess and B. Zumino, Nucl. Phys. B (Proc. Suppl.) {\bf 18B} 302
(1990).
\item B. Zumino, Mod. Phys. Lett. {\bf A} {\bf 13} 1225 (1991).
\item L. Hlavat\'{y}, J. Phys. A: Math. Gen. {\bf 25} 485 (1992).
\item W. Pusz and S.L. Woronowicz, Reports on Math. Phys. {\bf 27} 231 (1989).
\item B. Zumino, in: Unification of the Fundamental Particle Interactions (S.
Ferrara, J. Ellis and P. van Nieuwenhuizen Eds.),
Proc. Europhysics Conf., Erice (1980), Plenum Press 1980, p. 101.
\item W. Pusz, Reports on Math. Phys. {\bf 27} 349 (1989).
\item A. Schirrmacher, in: Groups and Related Topics
(R. Gielerak et al. Eds.), Kluwer Academic Publishers (1992)
p. 55.
\item A. Sudbery, York preprint PRINT-91-0498 (YORK).
\item A. Sudbery, York preprint YORK-92-1 .
\item G. Maltsiniotis, C. R. Acad. Sci. Paris, {\bf 331} 831 (1990).
\item G. Maltsiniotis, ``Calcul differentiel sur le groupe lin\'{e}aire
quantique'', ENS expos\'{e} (1990).
\item Yu. Manin, Bonn preprint MPI/91-47.
\item Yu. Manin, Bonn preprint MPI/91-60.
\item B. Zumino, in: Math. Phys. X (K. Schm\"udgen Ed.),
Proc. X-th IAMP Conf., Leipzig (1991), Springer-Verlag (1992), p.20.
\item N. Yu. Reshetikin, M. A. Semenov-Tian-Shansky, Lett. Math. Phys. {\bf 19}
133 (1990).
\item P.P. Kulish, R. Sasaki, C. Schwiebert, Preprint YITP/U-92-07.
\item P. Schupp, P. Watts and B. Zumino, Berkeley preprint, LBL-32315 and
UCB-PTH-92/14.
\item U. Carow-Watamura, M. Schlieker, S. Watamura, and W. Weich, Commun.
Math. Phys. {\bf 142} 605 (1991).
\item B. Drabant, B. Jur\v{c}o, M. Schlieker, W. Weich and B. Zumino, Preprint
MPI-PH-92-39, LBL-32354 and UCB-PTH 92/16.

\end{enumerate}

\end{document}